\documentclass[10pt,twocolumn]{article}
\usepackage{amsmath}
\usepackage{amsfonts}
\usepackage{amssymb}
\usepackage{graphicx}
\usepackage{bm}
\title{\textbf{Dissociative electron attachment dynamics of carbon disulfide and violation of axial recoil approximation near the 6-eV resonance}}
\author{\textbf{Anirban Paul$^{1\dag}$} and \textbf{Dhananjay Nandi$^{2\dag\ast}$}\\ $^\dag$Indian Institute of Science Education and Research Kolkata, Mohanpur 741246, India\\$^*$ Center for Atomic, Molecular and Optical Sciences $\&$ Technologies, \\Joint initiative of IIT Tirupati $\&$ IISER Tirupati, Yerpedu, 517619, Andhra Pradesh, India\\ \small{email:$^1$ap18rs006@iiserkol.ac.in, $^2$dhananjay@iiserkol.ac.in}}

\date{}
\begin{document}
\twocolumn[
  \begin{@twocolumnfalse}
    \maketitle
    \begin{abstract}
    Complete dissociation dynamics of low-energy electron attachment to carbon disulfide have been studied using the velocity slice imaging (VSI) technique. The ion yields of S$^-$ and CS$^-$ fragment ions as the function of incident electron energy in the range 5 to 11 eV have been obtained. Two resonances for S$^-$ ions at around 6.2 eV and 7.7 eV and only one resonance for CS$^-$ ions at around 6.2 eV have been obtained in this energy range. The kinetic energy and the angular distributions of these fragment negative ions at different incident electron energies around these resonances have been measured. From the angular distribution of these fragment anions, we have found that the bending of the temporary negative ions causes a significant change in the angular distribution from the expected one.
    \end{abstract}
  \end{@twocolumnfalse}
]

\section{Introduction}
Low energy electron-molecule collision is an important process in different branches of science. Radiation-induced damage of living cells is mainly caused by the low-energy secondary electrons produced by the high-energy primary beam. Low energy electron also plays a very significant role in the dissociation processes involved in the upper atmosphere. The low-energy electrons are also responsible for the depletion of the ozone layer. Recent studies have also found that the single- and double-strand break of DNA, etc. are primarily caused by these low-energy secondary electron impacts via the dissociative electron attachment (DEA) process\cite{Boudaiffa, Janusz, Pan2003}. Dissociative electron attachment (DEA) is a two-step resonant process and it is the dominant process in low energy inelastic electron collisions with the molecule. In DEA the incident electron is attached to the molecule and forms a temporary negative ion (TNI) and subsequently dissociates to a negative ion fragment and one or more than one neutral fragment. It is the dominant process for incident electron energy $\leq$ 20 eV. The functional group dependence of this process \cite{Prabhudesai2005} has made this process more interesting. To understand this process, it is important to perform similar experiments in the laboratory in a controlled way. \cite{Kawarai, Mahmoodi-Darian, Sahbani} The interstellar sulfur chemistry received a great thrust during the last few decades because of the detection of a series of sulfur-containing molecules in the interstellar medium and also in the cometary atmosphere \cite{Lovas2004} and 15 compounds have been detected in space till the date. Carbon disulfide is a sulfur-containing linear-triatomic molecule with carbon at the center with two sulfur connected to the carbon atom with two double bonds. It is a neurotoxic colorless volatile liquid at room temperature. It is often used as a building block in organic chemistry as well as an industrial and chemical non-polar solvent. DEA to CS$_2$ has been studied by Kraus, \cite{kraus1961bestimmung} Dillard and Franklin, \cite{Dillard} MacNeil and Thynna, \cite{macneil1969negative} Zeisel \textit{et al.}, \cite{Ziesel} Krishnakumar and Nagesha, \cite{Krishnakumar_1992} Nagesha \textit{et al.} \cite{nagesha1997theoretical} and Rangwala \textit{et al}. \cite{rangwala2001dissociative} The first three studies obtained the appearance energies and the peak positions of the different resonant peaks for different fragment negative ions from the DEA to CS$_2$. Zeisel $el$ $al.$ measured the absolute cross sections for the formation of negative ions but only in the energy range 2.7 to 4.3 eV, later Krishnakumar and Nagesha measured the absolute DEA cross-section for different negative ion production for DEA and also for ion-pair dissociation (IPD). Nagesha $et$ $al.$ gave the theoretical insight of production of S$_2^-$ ions from the DEA to CS$_2$. Rangwala and Krishnakumar studied the DEA process of electronically excited CS$_2$.\\
In this article, we have discussed the dissociation dynamics involved in the 6.2 eV resonance of CS$_2$. The kinetic energy along with the angular distribution of fragment S$^-$ and CS$^-$ ions have been extracted from the momentum image of the velocity slice imaging (VSI) spectrometer. While from the angular distribution of the fragment ions, we get information about the symmetry of the resonant states involved in the process. We have confined our discussion around the 6.2 eV resonance only and have measured the kinetic energy and angular distribution around this resonance.

\section{Experimental}
The whole experiment has been performed with the velocity slice imaging technique. The details of the technique have been discussed elsewhere.\cite{Eppink} The details of the experimental setup have also been reported earlier.\cite{nag, Nag_2015} We, therefore, have discussed the setup and the experimental procedure briefly here. The setup mainly consists of an electron gun, a Faraday cup, and a velocity slice imaging (VSI) spectrometer. The electron gun produces a pulsed (200 ns wide) electron beam which is further collimated magnetically by the magnetic field produced by two coils in the Helmholtz configuration. This collimated electron beam is made to cross at right angles with an effusive molecular beam produced by a capillary along the axis of the velocity slice imaging (VSI) spectrometer. 
The spectrometer consists of a pusher plate, a puller plate, a lens electrode, and a conical drift tube. The molecular beam is an effusive continuous one. The collision between the electron beam and molecular beam occurs at the interaction region, which is situated in the area between the pusher and puller plate, the effusive molecular beam enters the interaction region through a small hole in the middle of the pusher plate. 
A negative extraction pulse having a duration of 4 $\mu s$  is applied at the pusher plate after 100 ns of the electron gun pulse as the extraction field, to direct the newton sphere of fragment negative ions towards the detector, the puller plate is always remain grounded. The main principle \cite{Eppink} of the VSI spectrometer is to focus all the ions having the same velocity (speed + direction) on a single point of the detector. The lens plate is used for this focussing. Our spectrometer consists of one lens electrode to focus the ions. Following the lens plate, there is a conical-shaped drift tube, for our experimental purpose we have applied 108 V potential to the drift tube. The drift tube is used to expand the newton sphere of the fragment ions for better resolution. 
\begin{figure}[h]
\centering
  \includegraphics[scale=0.48]{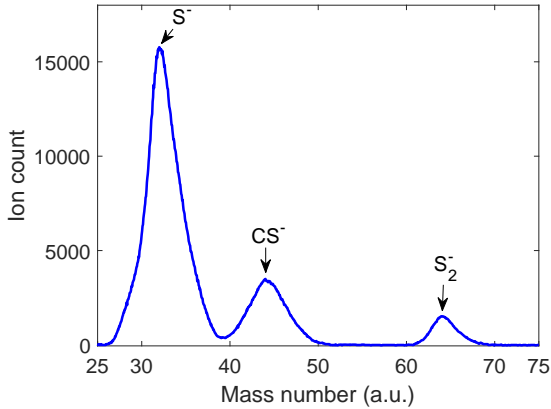}
  \caption{Mass spectra of various fragment anions produced from CS$_2$ due to the interaction of 6.2 eV incident electron energy.}
  \label{Mass_spectra}
\end{figure}
\begin{figure*}
 \centering
    \centerline{\includegraphics[scale=0.48]{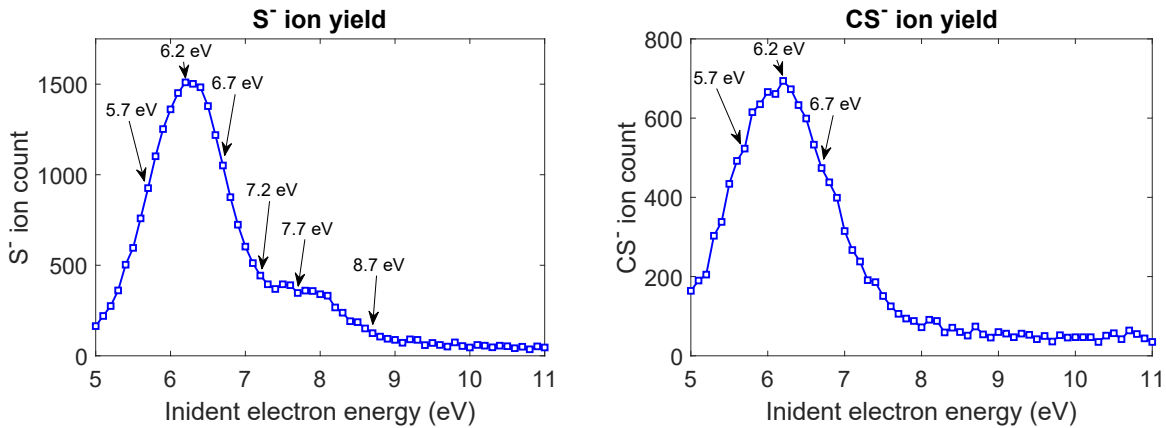}}
  \caption{\footnotesize{Ion yield curves of S$^-$ and CS$^-$ ions arising from DEA to CS$_2$. The arrows indicate the energies at which the velocity slice images are taken.}}
  \label{fig:ion_yield}
\end{figure*}
The VSI spectrometer is followed by a 2D position-sensitive detector (PSD) that consists of a set of three micro-channel plates mounted in a Z-stack configuration, and a position decoding delay line hexanode \cite{jagutzki2002multiple}, placed just after the MCP stack.
The time-of-flight (TOF) of the detected ions is determined from the back MCP signal and the x and y positions of each detected ion are collected by hexanodes. Thus, the x, y, and TOF of each ion can be collected and stored in a list-mode file format (.lmf), for our experimental purpose we have used the CoboldPC software from RoentDek to collect these data and store them in a .lmf file. Thus, we can construct the full newton sphere from the .lmf file. To construct the newton sphere from the data, we have to scale the TOF axis appropriately. For our analysis, we have used solid angle weighted slices.\cite{Moradmand2013} All the calibration is done by the Oxygen (O$_2$). The experiments have been performed using  99.9 $\%$ pure commercially available CS$_2$. 

\section{Results and Discussions}
The mass spectra of the negative ion fragments produced due to the DEA to CS$_2$ has been obtained for 6.2 eV incident electron energy as shown in Fig. \ref{Mass_spectra}. Three mass peaks correspond to S$^-$, CS$^-$ and S$_2^-$ are obtained. Previous studies \cite{Krishnakumar_1992,Ziesel,nagesha1997theoretical} also found C$^-$ ions at this resonance. However, we are unable to detect C$^{-}$ which could be due to the low cross-section for the formation, and the spectrometer is optimized for velocity map imaging condition. Here the interesting thing is the production of S$_2^-$ ions. For the production of S$_2^-$ ion from the DEA to CS$_2$, simultaneous breaking of two CS bonds and formation of a new S$-$S bond is required, and for that the two S atoms need to come closer by the bending to form an S$-$S bond. After the electron attachment in linear geometry, the resulting TNI may relax in some bent geometry. This type of bending of TNI states after the electron attachment for linear polyatomic molecules is evident from DEA dynamics of molecules like CO$_2$.\cite{Slaughter} On the other hand, the production of CS$^-$ ions required only one C$-$S bond-breaking while the production of S$^-$ ions is possible by the breaking of only one or both the C$-$S bonds. The bending of TNI after the electron attachment can affect the angular distribution of this S$ ^-$ and CS$^-$ fragment anions and thus deviate from the expected distribution.
The ion yields as a function of incident electron energy for different fragment negative ions have been obtained for the incident electron energy range from 5.0 to 11.0 eV.
The ion yields of S$^-$ and CS$^-$ fragment anions have been  shown in Figs. \ref{fig:ion_yield} (a) and \ref{fig:ion_yield} (b).
\begin{figure*}[ht]
 \centering
    \centerline{\includegraphics[scale=0.68]{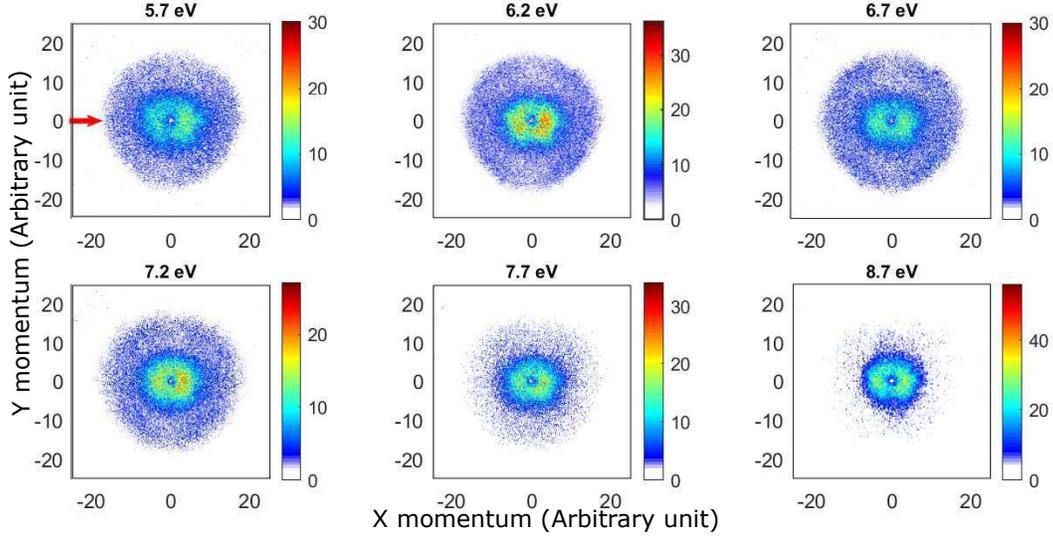}}
  \caption{\footnotesize{Sliced images of S$^-$/CS$_2$ ions at different electron energies. The electron beam direction is from left to right (The red arrow indicates the electron beam direction) through the center of each image. The radial size of the image represents the magnitude of the velocity of the ions}}
  \label{fig:vsi-images-S}
\end{figure*}
The ion yield curve of S$^-$ ions in \ref{fig:ion_yield} (a) shows a large resonance peaking at 6.2 eV with a smaller one at the falling tail of this bigger resonance at around 7.7 eV. Krishnakumar and Nagesha \cite{Krishnakumar_1992} found three resonances peaking at 6.2, 7.7, and 9.2 eV in this energy range. In the present work, we are not been able to detect the 9.2 eV peak, which is maybe because the cross-section is very low there and due to the relatively poor resolution of the electron source ($\sim$0.6 eV). \cite{nag}\\
The possible dissociation pathways/channels producing S$^-$ ions due to DEA are given below
\begin{align*}
e^-  +  CS_2 & \longrightarrow (CS_2^-)^* \tag{TNI formation}\\
(CS_2^-)^* & \longrightarrow
    \quad  S^-  + CS \tag{a} \\
     & \longrightarrow
    \quad  S + C + S^- \tag{b} \\
 	 & \longrightarrow       
    \quad  S^-  + CS \longrightarrow S^-  + C + S \tag{c} \\
  	 & \longrightarrow
    \quad  S  + CS^- \longrightarrow S + C + S^- \tag{d} 
\end{align*}
The first channel (channel (a)) among these four possible channels producing S$^-$ ions is a two-body dissociation channel, while all three other channels are three-body dissociation channels. Channels (b) is the symmetric bond dissociation channel, where, both the C$-$S bonds break simultaneously instead of being broken sequentially. While, channel (c) and channel (d) are the two sequential dissociation channels, where two C$-$S bonds break sequentially instead of being broken simultaneously. 
On the other hand, there is only one possible way for the production of the CS$^-$ ion, since, CS$^-$ ions can not be produced via a three-body dissociation channel,
\begin{equation*}
e^-  +  CS_2 \longrightarrow  (CS_2^-)^* \longrightarrow \quad  CS^- +  S\\
\end{equation*}
The thermodynamic threshold for different two-body DEA channels producing S$^-$ ions can be calculated theoretically using the following expression,
\begin{equation}
E_{Th}=D-A+E^*
\end{equation}
where $D$ is the CS$-$S bond dissociation energy (BDE), $A$ is the electron affinity of S, and $E^*$ is the energy (electronic + rovibrational) of the neutral CS fragment. If the CS fragment is in its electronic ground state then, $E^*$ = 0. Inserting the CS$-$S bond dissociation energy, $D$(CS$-$S) = 4.66 eV \cite{Ziesel} and electron affinity of S, $A$(S) = 2.08 eV \cite{nagesha1997theoretical} in equation (1), we get the calculated threshold value for the production of S$^-$ is 2.58 eV. On the other hand, the first electronic excitation energy of the CS radical is 4.82 eV. \cite{BRUNA1975} Therefore, the second-lowest threshold for the two-body dissociation channel is 7.40 eV.
For the three-body dissociation channels, on the other hand, breaking of both the C$-$S bonds is required. Therefore, the calculated threshold for these three-body dissociation processes is, $E_{Th}$ = (2$D - A$) = 7.24 eV. \\
For the production of S$^-$ ions from this sequential dissociation channel (d), the intermediate CS$^-$ must carry the E$_{int}$ value higher than the threshold energy (E$_{Th}$) of CS$^-$ $\rightarrow$ S$^-$ + C (4.66 eV). At the same time, if the intermediate CS$^-$ carries very high vibrational energy (E$_{int}$) which is higher than its electron affinity (EA(CS) = 1.66 eV) then it prefers to decay to the neutral CS radical by the electron autodetachment induced with molecular nuclear motions.\cite{Gao2020, Simons1981, Acharya1984} As a result, we can neglect the contribution of this channel (channel (d)) for the production of S$^-$.\\
The thermodynamic thresholds for the production of the CS$^-$ ions can be calculated using the following expression,
\begin{equation}
E_{Th}=D(CS-S)-A(CS)
\end{equation}
With the electron affinity of CS, $A$(CS) = 1.66 eV \cite{nagesha1997theoretical} the calculated threshold for the production of CS$^-$ becomes 3.0 eV.\\
The velocity slice images of S$^-$ ions are taken at six different incident electron energies around these two resonances and are shown in Fig. \ref{fig:vsi-images-S}. The black arrows in the ion-yield curve are the showing where these slices are taken. Up to 7.2 eV, we observe a blob in the sliced images along with a ring which indicates a low kinetic energy band along with a high energy band of S$^-$ ions. But beyond 7.2 eV, the outer ring starts to disappear. 
The presence of two distinct structures in the sliced images: the ring and the blob suggests that one low energy band and one high energy band of S$^-$ ions are present. From the sliced images four lobes can be obtained in the ring part. 
The presence of two distinct structures in the sliced images: the ring and the blob suggests that one low energy band and one high energy band of S$^-$ ions are present.
On the other hand, the sliced images of CS$^-$ ions are taken at three different incident electron energy around the only resonant peak and are shown in Fig. \ref{fig:vsi-images-CS}.
At 5.7 eV, we observe a blob in the momentum image indicating low kinetic energies of CS$^-$ ions. As we increase the electron energy to 6.7 eV, an outer ring appears but does not form a distinct ring. 

\subsection{Kinetic Energy Distribution}
\begin{figure}[ht]
\centering
  \includegraphics[scale=0.4]{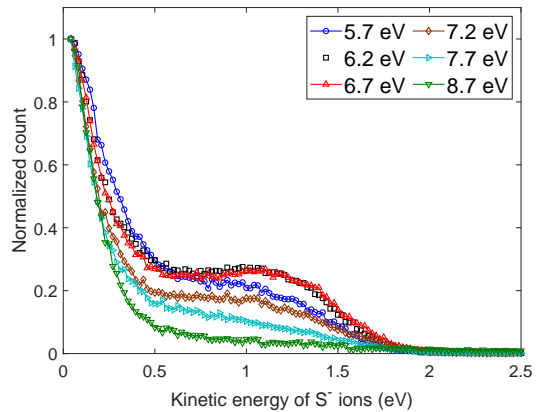}
  \caption{\footnotesize{Kinetic energy distributions of S$^-$ from
DEA to CS$_2$ at different incident electron energies. The distributions shown are obtained after integrating over entire $2\pi$ angles about the electron beam direction and normalized at the maximum value.}}
  \label{fig:KE_dist_S}
\end{figure}

\begin{figure*}[ht]
 \centering
    \centerline{\includegraphics[scale=0.68]{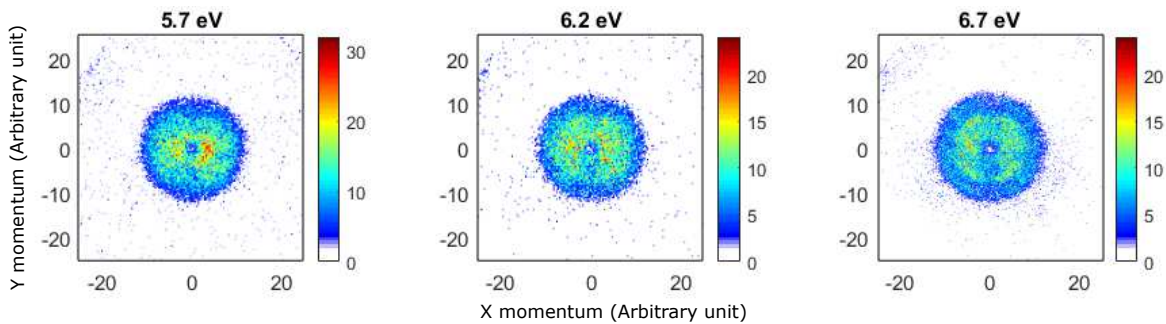}}
  \caption{\footnotesize{Wedge sliced images of CS$^-$/CS$_2$ ions at different electron energies. The electron beam direction is through the center of each image from left to right.}}
  \label{fig:vsi-images-CS}
\end{figure*}

The kinetic energy distributions of the S$^-$ ions, extracted from the above-sliced images are shown in Fig. \ref{fig:KE_dist_S}.
The kinetic energy distributions of the negative ion fragments are obtained by integrating over the entire ejection angle (2$\pi$ angle) of the fragment negative ions in the sliced data and plotting them as a function of the ion kinetic energy.
The present kinetic energy distributions are showing two peaks, one intense low energy peak with a peak position near zero and one higher energy peak peaking at around 1.2 eV. Nagesha \textit{et al}. \cite{nagesha1997theoretical} also reported the high energy ions at this resonance. The higher energy peak is a broad one and its peak position is not changing much with the incident electron energy, only the maximum kinetic energy is increasing slightly with the increase in incident electron energy.
The threshold calculation for different channels suggests that the second-lowest threshold is for the three-body dissociation channels having the threshold energy of 7.24 eV. Therefore, we can exclude all the other dissociation channels except the lowest energy dissociation channel (E$_{Th}$ = 2.58 eV) for incident electron energy $\leq$ 7.2 eV.
But, due to the finite resolution of our electron source, the abundance of the higher energy S$^-$ ions starts to decrease from 7.2 eV incident electron energy.
The threshold of the two-body dissociation channel producing S$^-$ is 2.58 eV, which leaves about 3.6 eV excess energy (at peak of the 6.2 eV resonance) to be distributed as translational and internal energy of the fragments. The larger peak in the distribution curve which is peaked at around 0 eV implies that the corresponding neutral CS fragment is in a highly excited vibrational ($\nu$) and/or rotational (j) state.
We have calculated the expected maximum kinetic energy of S$^-$ and CS$^-$ ions for different incident electron energies and listed them in table \ref{table:1}. From the experimentally obtained kinetic energy distribution of S$^-$ ions, we can see that the experimentally obtained maximum kinetic energies at different incident electron energies are less than the calculated maximum value at the corresponding electron energy which suggests that a part of the excess energy is distributed to the rotational and vibrational motion of the CS fragments and/or to the bending motion of the TNI. 
From the kinetic energy distribution, it can also be seen that the higher energy peak starts to disappear for incident electron energy of more than 7.2 eV. That is maybe because the small resonance at 7.7 eV mostly dissociates via the three-body dissociation channels ($E_{Th} = 7.24 $ eV) and/or via the second two-body dissociation channel ($E_{Th} = 7.40 $ eV).
\begin{figure}[ht]
\centering
  \includegraphics[scale=0.4]{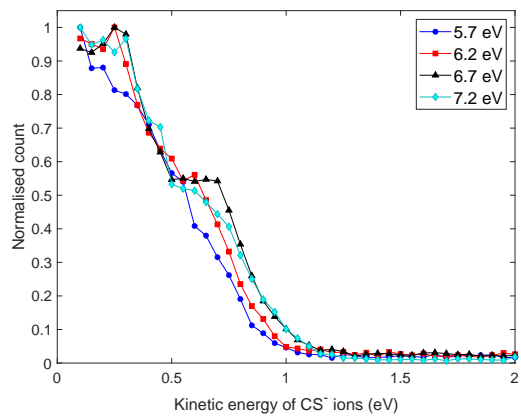}
  \caption{\footnotesize{Kinetic energy distributions of CS$^-$ from
DEA to CS$_2$ at different incident electron energies.}}
  \label{fig:KE_dist_CS}
\end{figure}
\begin{table}[h]
\centering
\begin{tabular}{|c|cc|}
\hline
Incident electron  & Max.$\quad$kinetic & energy$\quad$of \\
energy (eV) & S$^-$ ions (eV) &  CS$^-$ions (eV)\\
\hline
5.7  & 1.81  & 1.14\\
\hline
6.2  & 2.10  & 1.35\\
\hline
6.7  & 2.39  & 1.58\\
\hline
7.2  & 2.67  & 1.77\\
\hline
7.7  & 2.96  & -\\
\hline
8.7  & 3.54  & -\\
\hline
\end{tabular}
\caption{Incident electron energies and the expected maximum energies of the anions.}
\label{table:1}
\end{table}\\
The kinetic energy distribution of the CS$^-$ ions is plotted in Fig. \ref{fig:KE_dist_CS}, for different incident electron energies. The kinetic energy distributions are showing a relatively broader low energy peak, unlike that for the S$^-$ ions which have a sharp low energy peak, the higher energy CS$^-$ ions are also present there but not separable from the low energy peak. Here also can see that the experimentally obtained maximum kinetic energies at different incident electron energies are less than the calculated maximum value at the corresponding electron energy which suggests that a part of the excess energy is distributed to the rotational and vibrational motion of the CS$^-$ fragments and/or to the bending motion of the TNI.
The low energy peak in the kinetic energy distribution of CS$^-$ ions, on the other hand, can only be explained by the formation of highly rovibrationally excited CS$^-$ ions, since, the neutral S atom has its first excited state lies 25.78 eV higher\cite{duncan1952} than the ground state.

\subsection{Angular Distribution}
\begin{figure}[h!]
\centering
  \includegraphics[scale=0.42]{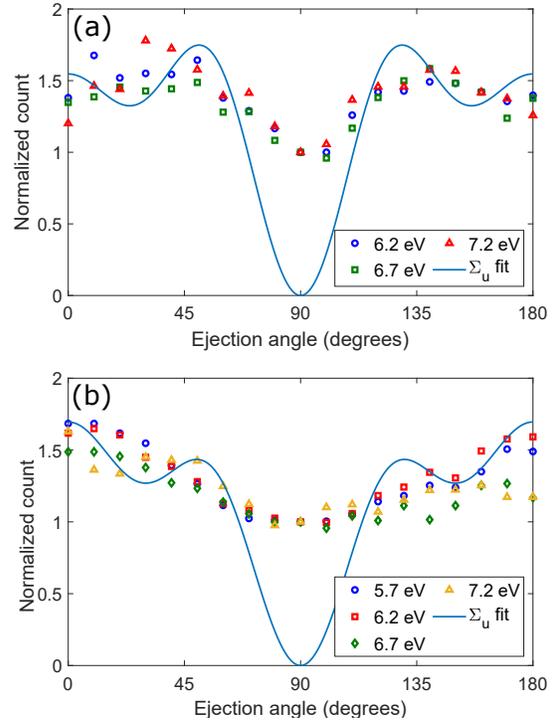}
  \caption{\footnotesize{(a) Angular distributions of the higher energy S$^-$ ions from the DEA to CS$_2$ at different incident electron energies. The distributions shown here are for ($\circ$) 6.2 eV, ($\square$) 6.7 eV, ($\bigtriangleup$) 7.2 eV incident electron energies, (b) Angular distributions of the lower energy S$^-$ ions from the DEA to CS$_2$ at different incident electron energies. The distributions shown here are for ($\circ$) 5.7 eV, ($\square$) 6.2 eV, ($\diamond$) 6.7 eV, ($\bigtriangleup$) and 7.2 eV incident electron energies. The solid curve in both these figures are the best fit for $\Sigma_u$ resonant symmetry considering axial recoil approximation to be valid.}}
  \label{fig:AD_All_S_Axial_recoil}
\end{figure}
\begin{figure}[h!]
\centering
  \includegraphics[scale=0.34]{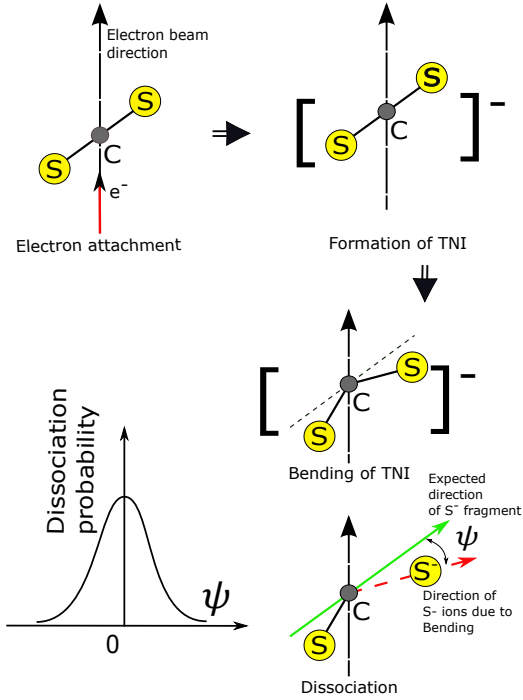}
  \caption{\footnotesize{The schematic explains the bending of the CS$_2$ molecule after electron attachment. The incident electron (the direction of the electron beam is shown with a black arrow through the C-atom) attaches to the molecule at its equilibrium geometry and forms a temporary negative ion (TNI) following the Franck-Condon transition rule. But after the attachment, the TNI may bend to get to its equilibrium bond angle in that resonant state.
  As a result, the C$-$S bond gets broken during the time of bending which causes deviation in the direction of the fragment anion trajectories. Here, in this figure, we have shown the ejection direction of the S$^-$ ion without the bending of the TNI by a red arrow while the green arrow indicates the ejection direction if the direction of the ejecting S$^-$ deviated by an angle $\psi$. Here we have considered the simplest model where we have considered that the dissociation occurs mostly at the 180$^\circ$ bond angle ($\psi$ = 0$^\circ$) and considered the C$-$S bond dissociation probability is a Gaussian function of deviation angle ($\psi$).}}
  \label{fig:TNI_Bending}
\end{figure}
The expression for the angular distribution of the fragment negative ions from the DEA to the diatomic molecule was first given by O'Malley and Taylor. \cite{O'Malley}
\begin{figure}[h!]
\centering
  \includegraphics[scale=0.42]{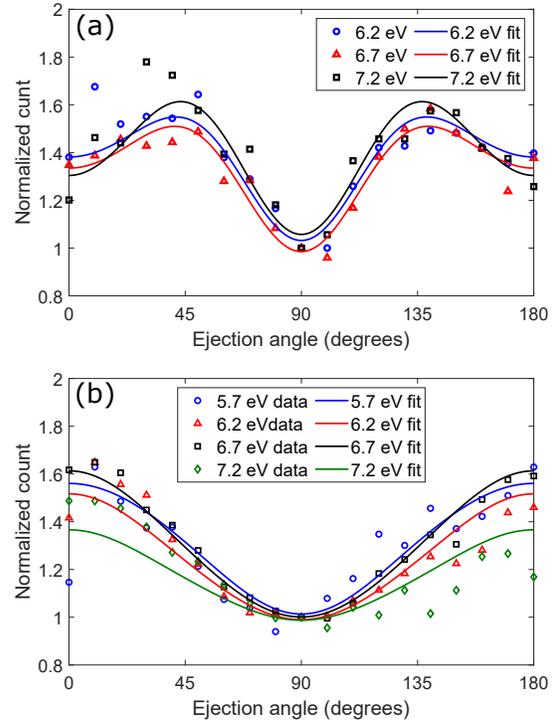}
  \caption{\footnotesize{(Color online) Fitting of the angular distribution of S$^-$ ions with $\Sigma_u $ symmetry considering bending of the molecule. (a) Fitting of the angular distribution of the high S$^-$ ions with $\Sigma_u $ symmetry considering bending of the molecule. (b) Fitting of the angular distribution of the low energy (near zero energy) S$^-$ ions with $\Sigma_u $ symmetry considering bending of the molecule. }}
  \label{fig:AD_All_S_with_bending}
\end{figure} 
The expression is as follows:
\begin{equation}
 I(\theta,\phi,k) = \sum_{ \mu } \vert \sum_{l = \mu } a_{l \mu}(k)Y_l^\mu(\theta,\phi)e^{i\delta_l}\vert^2   
\end{equation}
where $a_{l\mu}(k)$ are energy-dependent expansion coefficients, $k$ is the incident electron momentum, $Y_l^{\mu}(\theta,\phi)$ are the spherical harmonics, $\mu$ is the difference in the projection of the angular momentum along the inter-nuclear axis for the neutral molecular state and the negative ion resonant state, given as $\mu = \vert \Lambda_f - \Lambda_i \vert$, $l$ is the angular momentum of the incoming electron with values given by $l \geq \vert \mu \vert$ and ($\theta$, $\phi$) are the polar angles of the negative ion fragments with respect to the incident electron beam direction.\\
Later Azaria \textit{et al.} \cite{Azria} gave a general expression of the angular distribution of the negative ion fragments averaging over $\phi$ for polyatomic molecules. The expression is as follows:
\begin{eqnarray}
I(\theta) = \frac{1}{2 \pi}\int_{ 0 }^{2 \pi} \vert \sum_{l \mu \epsilon } i^l e^{i\delta_l} a_{l\mu}^{\epsilon} X_{l \mu}^{\epsilon}(\theta,\phi)\vert^2 d \phi
\end{eqnarray}
where $X_{l\mu}^{\epsilon}$ are the basis functions for the irreducible representation of the group of the molecule, $a_{l\mu}^{\epsilon}$ are their amplitude and all other variables are the same as discussed earlier.
CS$_2$ is a linear triatomic molecule and from our threshold calculation, we saw that the production of S$^-$ and CS$^-$ is due to the two-body dissociation and therefore, the S$-$CS bond-breaking can be considered similar to the diatomic bond-breaking case. Eqn. (3) therefore can be used to fit our angular distribution.
CS$_2$ is also centrosymmetric and as a result it has inversion symmetry and its ground state is $^1\Sigma^+_g $ and therefore $\Lambda_i$ = 0.\cite{nagesha1997theoretical} As a result, for $\Sigma_g$ to $\Sigma$, $\Pi$, $\Delta$ and $\Phi$ transitions $\mu$ = 0, 1, 2 and 3 respectively and since, it is centrosymmetric its states should also have garade (g) and ungarade (u) parity and the selection rules suggests that even $l$ values are responsible for garade (g) and odd $l$ values are responsible for ungarade (g) parity.

\subsubsection{Angular distribution of S$^-$ ions}
The angular distributions of the higher energy S$^-$ ions and the low energy S$^-$ ions are plotted in Figs. \ref{fig:AD_All_S_Axial_recoil} (a) and \ref{fig:AD_All_S_Axial_recoil} (b) respectively.
We have plotted the angular distributions of the higher energy band of the S$^-$ ions for incident electron energies 6.2, 6.7, and 7.2 eV respectively where the higher energy ions are prominent.
\begin{table*}[h!]
\small
  \caption{\ Fitting parameters for the angular distributions of the lower energy band of the S$^-$ ions.}
  \label{tbl:table2}
  \begin{tabular*}{\textwidth}{@{\extracolsep{\fill}}|llllll|}
  \hline
Electron energies &	$a_1$ &	$a_3$ &	Phase difference ($\delta$) & FWHM & R$^2$ value \\
\hline
 
5.7 eV  &	1.192 & 1.561 & 126.0$^\circ$ & 74$^\circ$ & 0.781\\

6.2 eV  &	0.965 &	1.793 & 129.5$^\circ$ & 74$^\circ$ & 0.923\\

6.7 eV  &	1.028 &	1.755 &	126.2$^\circ$ &	74$^\circ$ & 0.824\\

7.2 eV  &	0.894 &	1.836 &	136.0$^\circ$ &	74$^\circ$ & 0.810\\
 
\hline
  \end{tabular*}
\end{table*}
\begin{table*}[h!]
\small
  \caption{\ Fitting parameters for the angular distributions of the higher energy band of the S$^-$ ions.}
  \label{tbl:table3}
  \begin{tabular*}{\textwidth}{@{\extracolsep{\fill}}|llllll|}
  \hline
Electron energies &	$a_1$ &	$a_3$ &	Phase difference ($\delta$) & FWHM & R$^2$ value \\
\hline

6.2 eV  &	2.010 &	1.181 & 157.9$^\circ$ & 56$^\circ$ & 0.868\\

6.7 eV  &	1.901 &	1.270 &	145.7$^\circ$ &	56$^\circ$ &	0.927\\

7.2 eV  &	1.913 &	1.361 &	143.4$^\circ$ &	56$^\circ$ &	0.812\\
\hline 
  \end{tabular*}
\end{table*}  
On the other hand, we have plotted the angular distribution of the low energy (near-zero energy) S$^-$ ions for all the incident electron energies.
The angular distributions of the higher energy S$^-$ ions remain almost similar for different incident electron energies, the distributions are peaked at 40$^\circ$ and 140$^\circ$ angle with a slight forward-backward asymmetry with slightly more count in the forward direction. The reason for this asymmetry may be the population of bending and asymmetric mode of vibration of the target molecules.\cite{Slaughter}
Theoretical calculations by Nagesha \textit{et al}. found that the ground state of CS$_2^-$ have $^2\Pi_u$ symmetry with [$(^1\Sigma^+_g )$] $^5\pi_u^1$ configuration and a state having $(^2\Sigma^+_g )$ symmetry at 4.41 eV and a $(^2\Sigma^+_u )$ state at 6.15 eV.
We, therefore, have fitted the angular distributions of both the high and low energy S$^-$ ions with expression $ I(\theta) = \vert a_1 Y_1^1 + a_3 Y_3^1 e^{i\delta} \vert^2 $ which is the fitting expression for $\Sigma_g \rightarrow \Sigma_u $ transition with p and f partial waves in Figs. \ref{fig:AD_All_S_Axial_recoil} (a) and \ref{fig:AD_All_S_Axial_recoil} (b). The fitting is poor as can be observed from these figures.
Here, for our fitting, we have considered the axial recoil approximation to be valid. Axial recoil approximation is an approximation where the bond dissociation of the TNI occurs with the same geometry as it was during the time of electron attachment.
\begin{figure}[h!]
\centering
  \includegraphics[scale=0.42]{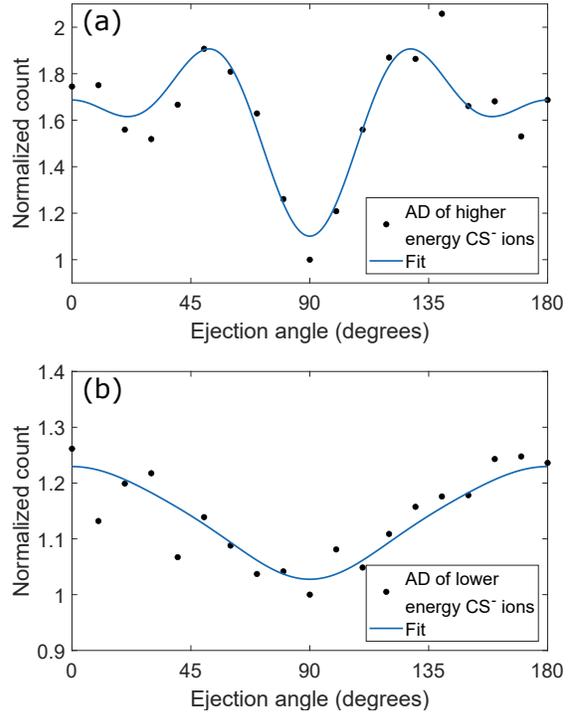}
  \caption{\footnotesize{(a) Fitting of the angular distribution of the high CS$^-$ ions with $\Sigma_u $ symmetry considering bending of the molecule. (b) Fitting of the angular distribution of the low energy (near zero energy) CS$^-$ ions with $\Sigma_u $ symmetry considering bending of the molecule. }}
  \label{fig:AD_CS_ions}
  \end{figure}
But as we have discussed earlier the bending of the TNI state due to the relaxation of the TNI is possible after the electron attachment. This type of bending of TNI can lead to the violation of axial recoil approximation. 
Since at room temperature only 1$\%$ of the CS$_2$ molecules can be in the v = 3 vibrationally excited state, which is the bending mode vibration\cite{Krishnakumar_1992} we can consider that the electron attachment to the CS$_2$ is occurring in its linear geometry and after the attachment the all these relaxations of TNI and dissociation dynamics are occurring.
Fig. \ref{fig:TNI_Bending} shows a schematic of such a process. The incident electron attaches to the molecule at its equilibrium geometry and forms a temporary negative ion (TNI) following the Franck-Condon transition rule. In the next step, the resultant TNI starts to bend to get to its equilibrium bond angle, and also due to the vibrational excitation of the TNI, one C$-$S bond may get broken during the time of bending which causes deviation in the direction of the fragment anion trajectories. In Fig. 8, we have shown the direction of the S$^-$ ion without the bending of the TNI by a red arrow while the green arrow indicates the direction if the dissociation occurs at a bond angle (180$^\circ$ $-$ 2$\psi$), $\psi$ represents the deviation. Here we have considered the simplest model where we have considered that the dissociation occurs mostly at the 180$^\circ$ bond angle ($\psi$ = 0$^\circ$) and considered the C$-$S bond dissociation probability is a Gaussian function of deviation angle ($\psi$) which suggests the TNI dissociates mostly at linear geometry but a significant contribution to the dissociation from the bent structure is present. 
\begin{table*}[h!]
\small
  \caption{\ Fitting parameters for the angular distributions of the CS$^-$ ions for 6.2 eV incident electron energy.}
  \label{tbl:table4}
  \begin{tabular*}{\textwidth}{@{\extracolsep{\fill}}|llllll|}
  \hline
AD of CS$^-$ ions in &	$a_1$ &	$a_3$ &	Phase difference ($\delta$) & FWHM & R$^2$ value \\
\hline

Lower energy band  &	1.980 &	1.537 & 140.3$^\circ$ & 48$^\circ$ & 0.876\\

Higher energy band  &	1.124 &	0.206 &	172.3$^\circ$ &	74$^\circ$ &	0.779\\

\hline 
  \end{tabular*}
\end{table*}
The angular distribution of the fragment anions due to the bending of the TNI can therefore be found by convolving the angular distribution with this Gaussian function.
\begin{equation}
    I_{Bending}(\theta) = \int_{-\pi}^{\pi} e^{-\frac{\psi^2}{2\sigma^2}}\,I_{Axial \, recoil}(\theta - \psi) \,d\psi
\end{equation}
Where, $ I_{Axial \, recoil}(\theta) = \vert a_1 Y_1^1 + a_3 Y_3^1 e^{i\delta} \vert^2 $.
The angular distribution fitted with Eqn. (5) is shown in Figs. \ref{fig:AD_All_S_with_bending} (a) and \ref{fig:AD_All_S_with_bending} (b) for both the higher and the lower energy ions. The fitting parameters for these fits are listed in rables \ref{tbl:table2} and \ref{tbl:table3}. From these fitting parameters, we can see that for high energy ions the FWHM of the Gaussian function used for the convolution is 56$^\circ$ while that is for low energy ions is 74$^\circ$. As we know that with the increase in FWHM value the Gaussian function gets flattered which means a relatively higher contribution from the bent structure of the TNI. A similar result was previously found by Ram \textit{et al.} in DEA to water.\cite{Ram}\\ 
There is no past theoretical calculation to suggest which resonance symmetry is involved in the 7.7 eV resonance.
To get more insights into the dynamics involved in this resonance, sophisticated dynamic simulations of the nuclear motions should be theoretically performed on the multi-dimensional potential energy surface of the TNI state. However, due to the unaffordable computations, they are currently unavailable.
Since, as stated earlier that the violation of axial recoil approximation is much for low energy ions we can not conclude much about the symmetry of the resonance from these angular distributions.

\subsubsection{Angular distribution of CS$^-$ ions}
Although the high energy ring is not separable from the inner blob as evident from the kinetic energy distribution curves of CS$^-$ ions, we have plotted the angular distribution of the CS$^-$ ions for ions having energy $\geq$ 0.5 eV and for ions having energy from 0 to 0.4 eV for 6.7 eV incident electron energy in Figs. \ref{fig:AD_CS_ions} (a) and (b) respectively.
The angular distribution of CS$^-$ is similar to that of the S$^-$ ions, the only difference is that it is showing a more backward count.
The angular distribution of the CS$^-$ ions in the higher energy band (having energy $\geq$ 0.5) are fitted with the expression given in Eqn. (5) with for the Gaussian function having FWHM of 48$^\circ$ while that is for the ions in the lower energy band(0 to 0.4 eV) ions is 74$^\circ$. 
The angular distributions of the CS$^-$ ions are almost similar to that of the S$^-$ ions at this 6.2 eV resonance.

\section{Conclusions}
The ion yield of S$^-$ ions shows a big resonance peaking at around 6.2 eV with a smaller one at around 7.7 eV, while the ion yield of CS$^-$ ions shows only one resonance peaking at around 6.2 eV. 
The kinetic energy distribution of S$^-$ ions shows two distinct peaks. The low energy peak is near 0 eV with its peak position not changing with the incident electron energy, which suggests that the low energy S$^-$ ions are due to the CS fragment excited to some very highly vibrationally excited state/states. 
The intensity of the high-energy band starts to decrease beyond 7.2 eV incident electron energy. This may be because the TNIs at the 7.7 eV resonance dissociate via the three-body dissociation channels (threshold energy = 7.24 eV) and/or through the second two-body dissociation channel (threshold energy = 7.40 eV).
The kinetic energy distributions of CS$^-$ ions is also consisting of two peaks where the low energy peak suggests that the CS$^-$ ions are in highly vibrationally excited states.
The experimentally obtained angular distributions of the negative ion fragments cannot be justified by the axial recoil approximation. The angular distribution can be justified by the bending of the molecule and the violation of axial recoil approximation.

\section*{Acknowledgements}
A. P. deeply appreciates the ``Council of Scientific and Industrial Research (CSIR)'' for the financial assistance. D.N. gratefully acknowledges the financial support from the ``Science and Engineering Research Board (SERB)'' under Project No. ``CRG/2019/000872''.

\bibliographystyle{h-physrev}
\bibliography{main_2.bib}

\begin{thebibliography}{10}

\bibitem{Boudaiffa}
B.~Badia, C.~Pierre, H.~Darel, A.~H. Michael, and S.~Léon,
\newblock Science {\bf 287}, 1658 (2000).

\bibitem{Janusz}
J.~Rak {\em et~al.},
\newblock {\em Stable Valence Anions of Nucleic Acid Bases and DNA Strand
  Breaks Induced by Low Energy Electrons} (, 2008), pp. 619--667.

\bibitem{Pan2003}
X.~Pan, P.~Cloutier, D.~Hunting, and L.~Sanche,
\newblock Phys. Rev. Lett. {\bf 90}, 208102 (2003).

\bibitem{Prabhudesai2005}
V.~S. Prabhudesai, A.~H. Kelkar, D.~Nandi, and E.~Krishnakumar,
\newblock Phys. Rev. Lett. {\bf 95}, 143202 (2005).

\bibitem{Kawarai}
Y.~Kawarai {\em et~al.},
\newblock The Journal of Physical Chemistry Letters {\bf 5}, 3854 (2014).

\bibitem{Mahmoodi-Darian}
M.~Mahmoodi-Darian {\em et~al.},
\newblock International Journal of Mass Spectrometry {\bf 293}, 51  (2010).

\bibitem{Sahbani}
S.~Kouass~Sahbani, P.~Cloutier, A.~D. Bass, D.~J. Hunting, and L.~Sanche,
\newblock The Journal of Physical Chemistry Letters {\bf 6}, 3911 (2015).

\bibitem{Lovas2004}
F.~J. Lovas,
\newblock Journal of Physical and Chemical Reference Data {\bf 33}, 177 (2004),
  https://doi.org/10.1063/1.1633275.

\bibitem{kraus1961bestimmung}
K.~Kraus,
\newblock Zeitschrift f{\"u}r Naturforschung A {\bf 16}, 1378 (1961).

\bibitem{Dillard}
J.~G. Dillard and J.~L. Franklin,
\newblock The Journal of Chemical Physics {\bf 48}, 2349 (1968),
  https://doi.org/10.1063/1.1669435.

\bibitem{macneil1969negative}
K.~MacNeil and J.~C. Thynne,
\newblock The Journal of Physical Chemistry {\bf 73}, 2960 (1969).

\bibitem{Ziesel}
J.~P. Ziesel, G.~J. Schulz, and J.~Milhaud,
\newblock The Journal of Chemical Physics {\bf 62}, 1936 (1975),
  https://doi.org/10.1063/1.430681.

\bibitem{Krishnakumar_1992}
E.~Krishnakumar and K.~Nagesha,
\newblock Journal of Physics B: Atomic, Molecular and Optical Physics {\bf 25},
  1645 (1992).

\bibitem{nagesha1997theoretical}
K.~Nagesha, B.~Bapat, V.~Marathe, and E.~Krishnakumar,
\newblock Zeitschrift f{\"u}r Physik D Atoms, Molecules and Clusters {\bf 41},
  261 (1997).

\bibitem{rangwala2001dissociative}
S.~Rangwala, S.~Kumar, and E.~Krishnakumar,
\newblock Physical Review A {\bf 64}, 012707 (2001).

\bibitem{Eppink}
A.~T. J.~B. Eppink and D.~H. Parker,
\newblock Review of Scientific Instruments {\bf 68}, 3477 (1997).

\bibitem{nag}
P.~Nag and D.~Nandi,
\newblock The European Physical Journal D {\bf 72}, 25 (2018).

\bibitem{Nag_2015}
P.~Nag and D.~Nandi,
\newblock Measurement Science and Technology {\bf 26}, 095007 (2015).

\bibitem{jagutzki2002multiple}
O.~Jagutzki {\em et~al.},
\newblock IEEE Transactions on Nuclear Science {\bf 49}, 2477 (2002).

\bibitem{Moradmand2013}
A.~Moradmand {\em et~al.},
\newblock Phys. Rev. A {\bf 88}, 032703 (2013).

\bibitem{Slaughter}
D.~S. Slaughter {\em et~al.},
\newblock  {\bf 44}, 205203 (2011).

\bibitem{BRUNA1975}
P.~Bruna, W.~Kammer, and K.~Vasudevan,
\newblock Chemical Physics {\bf 9}, 91 (1975).

\bibitem{Gao2020}
X.-F. Gao, H.~Li, X.~Meng, J.-C. Xie, and S.~X. Tian,
\newblock The Journal of Chemical Physics {\bf 152}, 084305 (2020),
  https://doi.org/10.1063/1.5135609.

\bibitem{Simons1981}
J.~Simons,
\newblock Journal of the American Chemical Society {\bf 103}, 3971 (1981),
  https://doi.org/10.1021/ja00404a002.

\bibitem{Acharya1984}
P.~K. Acharya, R.~A. Kendall, and J.~Simons,
\newblock Journal of the American Chemical Society {\bf 106}, 3402 (1984),
  https://doi.org/10.1021/ja00324a003.

\bibitem{duncan1952}
A.~B.~F. Duncan,
\newblock The Journal of Chemical Physics {\bf 20}, 951 (1952),
  https://doi.org/10.1063/1.1700656.

\bibitem{O'Malley}
T.~F. O'Malley and H.~S. Taylor,
\newblock Phys. Rev. {\bf 176}, 207 (1968).

\bibitem{Azria}
R.~Azria, Y.~L. Coat, G.~Lefevre, and D.~Simon,
\newblock Journal of Physics B: Atomic and Molecular Physics {\bf 12}, 679
  (1979).

\bibitem{Ram}
N.~B. RAM, V.~S. PRABHUDESAI, and E.~KRISHNAKUMAR,
\newblock Journal of Chemical Sciences {\bf 124}, 271 (2012).

\end{thebibliography}

\end{document}